\documentstyle{article}

\def \bi{\bibitem}
\def \Tr{{\rm Tr}}

\begin{document}

\title{Against Chaos in Temperature \\in Mean Field Spin Glass Models}
\author{
Tommaso Rizzo\\
\small Dipartimento
di Scienze Fisiche, Universit\`a ``Federico II'',\\
\small Complesso Monte S. Angelo, I--80126 Napoli (Italy)\\
\small e-mail: {\tt me2969@mclink.it}}
\date{March 2001}
\maketitle

\begin{abstract}
We study the problem of chaos in temperature in some mean-field
spin-glass models by means of a replica computation over a model
of coupled systems. We propose a set of solutions of the
saddle point equations which are intrinsically non-chaotic and
solve a general problem regarding the consistency of their
structure. These solutions are relevant in the case of
uncoupled systems too, therefore they imply a non-trivial overlap distribution $P(q_{T1T2})$ between systems at different temperatures. The existence of such solutions is
checked to fifth order in an expansion near the critical
temperature through highly non-trivial cancellations, while it is proved that a dangerous set of such cancellations holds exactly at
all orders in the Sherrington-Kirkpatrick (SK) model. The SK model with soft-spin distribution is also considered obtaining analogous results. Previous analytical results
are discussed.
\end{abstract}

\section{Introduction}

In this paper we shall address by analytical means the problem of
the correlations between the equilibrium states at different
temperatures of the Sherrington-Kirkpatrick spin-glass model. At any
temperature below the critical one there are infinitely many pure
states defined by the local values of the magnetization. The
correlation between two states are measured by the overlap
$q_{\alpha \beta}=\sum_i m_i^{\alpha}m_1^{\beta}$ wich according
to the Parisi solution can take values between zero and some
$q_{EA}$ which is the self-overlap of the states \cite{MPV}.

While pure states at a given temperature are correlated, for many years it was believed that states at different
temperatures were completely uncorrelated: this is the hypothesis of chaos in temperature. It was originally phrased as a constitutive ingredient of the
phenomenological droplet theory\ \cite{BM87,FH}, because otherwise
the growth of domains of correlated phases would give strong cooling
rate dependence and would not exibit the so-called rejuvenation effect. Contrary to this real spin glasses are to a large extent insensitive to the
cooling rate; furthermore, if we let an observable relax at a
given temperature in the glassy phase and then lower the
temperature we observe that its value skips to a higher value and
the system apparently rejuvenates. So everything seems to happen as
if there were great changes in the free-energy landscape, i.e.
chaos in temperature. 

A few years after the first
observations of this phenomenon \cite{LSB83}, it was discovered
that when the sample is heated back to the original
temperature the observable resumes the value it had immediately
before the second quench \cite{MEeCH}; therefore while the first
effect suggests that the information belonging to aging at a
higher temperature is destroyed upon cooling the system (chaos
effect), the second one implies no information loss at all
(memory effect). The two effects are apparently contradictory and their
explanation, particularly from a real-space point of view,
attracts large interest \cite{Misc1,Misc2,Misc3}. 

At the mean-field level it was shown that the equations of dynamics capture the physics of the two phenomena \cite{CK}. 

From a phase-space point of
view it seems rather difficult that a purely chaotic picture of
the temperature evolution of the free-energy landscape could
account for both chaos and memory effects and this is the main
reason for our analysis of the equilibrium states correlations.
 We must be careful on this point since the theoretical work of 
 the last ten years has raised many questions on the possibility 
  of explaining off-equilibrium dynamics by means of the static mean-field free-energy landscape \cite{CKBM}. 
At any rate, the ideas
  arising from the Parisi solution of the mean-field Sherrington-Kirkpatrick
  (SK) model have often proved to be very fruitful, for example
  in constructing phase-space pictures of dynamics like the 
  traps model \cite{BOU}. In particular a  scenario which had been suggested as soon as the
  ultrametric organization of the states was discovered \cite{MPNTV}
  was advocated in order to explain the rejuvenation and memory effects
  \cite{Met}. This scenario deals with the idea that each valley in the free-energy
  landscape bifurcates into many others when the temperature is
  lowered, so that rejuvenation is accounted for by equilibration
  between the newly born valleys, while memory is due to the fact
  that the topological structure of the states tree is preserved.
  The explanation of the two effects in this picture is completely
  different from the chaotic one and is supported by the fact that
  both rejuvenation and memory have been observed in numerical
  simulations on an intrinsically non-chaotic model like the GREM\
  \cite{grem}.

Our concern here is confined to check the existence of
correlations between equilibrium states at different temperatures.
We do not purport to give a full description of the structure, if any, of
such correlations but merely to collect evidence that they do
exist in contrast with the chaos hypothesis. 

Chaos is known to hold in mean-field models with different
magnetic fields, while the case of equal temperatures has been
treated by Sompolinsky in an unpublished work cited by Binder and
Young in their 1986 review \cite{BY}. In that context it is claimed that
there is chaos in temperature in mean-field spin glasses; we
don't agree with these findings as we shall explain at the end of
section 2. Kondor in 1989 addressed a related problem, i.e. the
spatial correlations between different temperatures states in
finite-dimensional spin glasses. Since chaos was generally accepted
at the time he was quite surprised in finding such correlations
to be infinitely long-ranged at zero-loop order \cite{Kondor1}, a result that is
deeply connected to our own. However, later Kondor and V\'egs\"o
showed that at one-loop order the correlation length becomes
finite \cite{Kondor2}.

The problem has also been studied by means of numerical simulations 
\cite{Rit,Ney,FN};
  our findings are in agreement with the recent work of Marinari and Billoire
\cite{MarBill}.
The existence of correlations at different temperatures is well shown numerically, the problem is rather that rejuvenation is not seen in simulations of realistic spin-glass models \cite{FelFed}. Correlations between states at different temperatures have been recently studied also within the TAP approach \cite{PPM}

\vskip5pt
The paper is organized as follows. In the next section we shall
present the model of coupled replicas of a SK spin-glass and discuss its
relevance to the problem of chaos. In section 3 we shall write
down the saddle-point equations and propose a set of solutions with a certain structure.
The validity of such solutions has been checked perturbatively to fifth order in the
order parameter and it turned out that they are non-chaotic i.e. their free energy at this order is the same as in the uncoupled case. We prove that at all orders these solutions, if they exist, are non-chaotic, showing that this property relies on their structure.
Therefore the problem is to check wether the SP equations can be solved at all orders with solutions of these kind. To this extent we provide in section 4 an exact result concerning a particular set of cancellations in the SP equations showing that they are verified at all orders. This result is connected with Kondor zero-loop expansion in replica field theory. 
However the possibility of solving the SP equations with the solutions we propose relies also on  others concellations, and we are not able to prove that {\em all} these cancellations actually hold at all orders.

Since the problem is to check wether the solutions we propose actually exist at all orders in section 5 we investigate a mechanism that could led to a breakdown of the scheme at higher orders.
The essential features of the problem can be explained starting from the SP equations. As we will see in section 3, the SP equations involve three $n\times n$ matrices $Q_1$,$Q_2$ and $P$ that will be parametrized in the standard Parisi form. For instance, the SP equation obtained differentiating the free-energy functional with respect to $Q_1$ has the form
\begin{equation}
\tau_1Q_{1ab}-B_{ab}(Q_1,Q_2,P)=0\label{SPint}
\end{equation}
Now, a necessary condition to solve the previous equation is that the second term $B$ must have the same structure of the first term $Q_1$, this is what we call {\em structural consistency}. Translated in the language of Parisi matrices, it means that if the function $q_1(x)$ has a plateau for $x$ greater that some $x_{1max}$ the function $b(x)$ must have a plateau for $x$ greater than the same $x_{1max}$.
For the standard Parisi solution it is easily seen that this is true because $B$ is a function only of $Q_1$, but in our case $B$ depends also on $Q_2$ and $P$.
If the functions $q_1(x)$,$q_2(x)$ and $p(x)$ have different structures, i.e. they display plateaus of unequal lenght located at different positions, as in our case, it is not trivial that $b(Q_1,Q_2,P)(x)$ has the same structure of $q_1(x)$.
We will see that in general this is not true, however we will prove that when the three functions $q_1(x)$,$q_2(x)$ and $p(x)$ has the structure we propose for them the function $b(x)$ has the same structure of $q_1(x)$, i.e. our solutions fullfill the necessarity condition to solve the previous SP equation.

This result is very general, it prevents the solutions we propose to break down due to structural inconsistencies. However it only states that $B_{ab}(Q_1,Q_2,P)$ has the same structure of $Q_1$, but doesn't say if it's really possible to tune the three functions $q_1(x)$,$q_2(x)$ and $p(x)$ in such a way that $B_{ab}(Q_1,Q_2,P)$ is exactly equal to $\tau_1Q_{1ab}$ in order to solve the SP equation (\ref{SPint}).
The arguments provided in section 4 and 5 are two necessary conditions for the existence of non-chaotic solutions; they are completely independent since the first is quantitative while the second is qualitative, however none of them is sufficient to prove that the solutions actually exist at all orders.

In section 6
the generalization of the SK model with soft spins will be studied evidencing again absence
of chaos. In section 7 we give our conclusions. The solutions are
reported in the appendix.

\section{The Model}

We consider a system composed of two replicas of a
Sherrington-Kirkpatrick (SK) spin glass model constrained to have
fixed values of their mutual overlaps. This model was first
studied in the case of replicas of the same temperature \cite{FPV1}; here we shall use the generalization to replicas with
two different temperature below the critical one \cite{FN}.
Denoting by $S_{i}^{r}$ the i-th spin of the r-th replica, we fix
a constraint
\begin{equation}
q_{c}=\frac{1}{N}\sum_{i=1}^{N} S_{i}^{1}S_{i}^{2}
\label{constraint}
\end{equation}
where $N$ is the total number of spins. Including the temperature
difference, the Hamiltonian of the system reads
\begin{equation}
H=-\sum_{i<j}J_{ij}(\beta_{1}S_{i}^{1}S_{j}^{1}+\beta_{2}S_{i}^{2}S_{j}^{2})
\end{equation}
So we take the same realization of the quenched
$\{J_{ij}\}$ for the two systems. They are chosen with Gaussian probability, zero mean
and variance $\frac{1}{N}$. The partition function is restricted
to those spin configurations that satisfy (\ref{constraint}). The
constraint (\ref{constraint}) is implemented introducing a
Lagrange multiplier $\epsilon$
\begin{equation}
Z=\sum_{\{S_{i}^{1},S_{i}^{2}\}}\int_{-i\infty}^{i\infty}\frac{d\epsilon}{2\pi}\
\exp \ \left[-H-\epsilon\ \left(\sum_{i=1}^{N}
S_{i}^{1}S_{i}^{2}-N\/q_{c}\right)\right]
\end{equation}
Instead of fixing the constraint $q_{c}$ we can consider the
partition function corresponding to the following Hamiltonian
\begin{equation}
H(\epsilon)=-\sum_{i<j}J_{ij}(\beta_{1}S_{i}^{1}S_{j}^{1}+\beta_{2}S_{i}^{2}S_{j}^{2})-\epsilon\
\sum_{i=1}^{N} S_{i}^{1}S_{i}^{2}
\end{equation}
this corresponds to systems coupled by a forcing term which
selects configurations with higher overlap. In the thermodynamic
limit the two description are obtained one from the other by a
Legendre transformation, in particular defining $F=-\ln Z$ the
following relation hold:
\begin{equation}
\label{legtrasf}
\epsilon\ = \frac{\partial{F(q_{c})}}{\partial{q_{c}}}
\end{equation}
Introducing replicas to average over the disorder we obtain via
standard manipulation the average partition function to the power
{\em n}
\begin{eqnarray}
 \overline{Z^{n}}=S.P.\ \exp\
\left[\frac{N}{4}\beta_{1}^{2}\Tr Q_{1}^{2}+\frac{N}{4}\beta_{2}^{2}\Tr
Q_{2}^{2}+\frac{N}{2}\beta_{1} \beta_{2}\Tr P^{2}\right. \nonumber
\\
\label{F}
\left. -N \ln Z[\hat{Q}]-N
q_{c}\sum_{\alpha}\epsilon_{\alpha}-\frac{N}{2}\sum_{\alpha}(\frac{\epsilon_{\alpha}^{2}}
{\beta_{1}\beta_{2}}-2P_{\alpha\alpha}\epsilon_{\alpha})\right]
\\
Z[\hat{Q}]=\sum_{\{S^{1}_{\alpha},S^{2}_{\alpha}\}} \exp\
\left[\frac{1}{2}\beta_{1}^{2}\sum_{\alpha\beta}Q_{1\alpha\beta}S^{1}_{\alpha}S^{1}_{\beta}\right.
\nonumber
\\
\left.+\frac{1}{2}\beta_{2}^{2}\sum_{\alpha\beta}Q_{2\alpha\beta}S^{2}_{\alpha}
S^{2}_{\beta}+\beta_{1}\beta_{2}\sum_{\alpha\beta}P_{\alpha\beta}S^{1}_{\alpha}S^{2}_{\beta}
\right]
\end{eqnarray}
where by S.P. we mean the value computed at the saddle point with
respect to the set $\{\epsilon_{\alpha}\}$ and
 to the order parameter which is a ${\em 2n \times\ 2n}$ matrix
 $\hat{Q}=\left(\begin{array}{cc}
 Q_{1} & P \\
 P^{t} & Q_{2}
 \end{array}\right)$.
The saddle-point (SP) equations then read
\begin{eqnarray}
Q_{1\alpha\beta}=\langle
S^{1}_{\alpha}S^{1}_{\beta}\rangle&Q_{2\alpha\beta}=\langle
S^{2}_{\alpha}S^{2}_{\beta}\rangle \nonumber
\\
\label{SPeq} P_{\alpha\beta}=\langle
S^{1}_{\alpha}S^{2}_{\beta}\rangle\
+\frac{1}{\beta_{1}\beta_{2}}\/\epsilon_{\alpha}\delta_{\alpha\beta}
&\epsilon_{\alpha}=\beta_1\beta_2(P_{\alpha\alpha}-q_{c})
\end{eqnarray}
Where the square brackets mean average taken with respect to the
Hamiltonian
\begin{eqnarray}
H=\frac{1}{2}\beta_{1}^{2}\sum_{\alpha\beta}Q_{1\alpha\beta}S^{1}_{\alpha}S^{1}_{\beta}
+\frac{1}{2}\beta_{2}^{2}\sum_{\alpha\beta}Q_{2\alpha\beta}S^{2}_{\alpha}S^{2}_{\beta}
+\beta_{1}\beta_{2}\sum_{\alpha\beta}P_{\alpha\beta}S^{1}_{\alpha}S^{2}_{\beta}
\end{eqnarray}
The SP equation in the unconstrained case can be
obtained setting $\epsilon$ to zero and neglecting the last
equation in (\ref{SPeq}). We eliminate the Lagrange multipliers
$\{\epsilon_{\alpha}\}$ in (\ref{F}) replacing their saddle point
values. This give for the $\epsilon$-dependent term in
$\overline{Z^{n}}$
\begin{equation}
-N
q_{c}\sum_{\alpha}\epsilon_{\alpha}-\frac{N}{2}\sum_{\alpha}(\frac{\epsilon_{\alpha}^{2}}{\beta_{1}\beta_{2}}-2P_{\alpha\alpha}\epsilon_{\alpha})=N\frac{\beta_{1}\beta_{2}}{2}\
\sum_{\alpha}(P_{\alpha\alpha}-q_{c})^{2}
\end{equation}
To solve the model we need a variational ansatz for the matrices
$Q_{1},Q_{2},$ and $P$. We choose each of them to be a Parisi
hierarchical matrix, in particular this fixes $P_{\alpha\alpha}=p_d$ and $\epsilon_{\alpha}=\epsilon$ for any replica index $\alpha$.

Let us comment the relevance of this model to the problem of
chaos. The mathematical formulation of the chaos hypothesis is
\begin{equation}
\overline{\langle S_i\rangle _{T1}\langle S_i\rangle _{T2}}=0
\end{equation}
That is, any two equilibrium states at different temperatures have zero overlap. Now, absence of chaos would imply that there
are pure states at different temperatures with non-zero overlap, so
that we would have a non-trivial function $P(q_{T1T2})$ that
measures the probability of finding an overlap $q$ between two
pure states of systems at different temperatures weighed
according to the Gibbs measure. Turning to our model, we see that
if we choose the constraint inside the support of the function
$P(q_{T1T2})$ the only effect will be to select those couples of
pure states that satisfy the constraining relation. As the number
of pure states grows less than exponentially with $N$, selection of 
only some couples of pure states instead of all will not change
the free energy or other extensive quantities. So in the absence of 
chaos we expect to find for some values of the constraint the same
values of the free energy as in the unconstrained case (i.e. the sum of the free energies of the two uncoupled systems)
while in the presence of chaos the free energy could increase.

 So far we have established that an increase in free
energy implies chaos, but what if we find no such an increase? Let
us show that this would imply a nontrivial $P(q_{T1T2})$, i.e.
absence of chaos. Suppose we find a set of solutions with the same
free energy of the unconstrained case corresponding to values
of the constraint $q_{c}$ spanning an open set from zero to some $p_{max}$; then from
(\ref{legtrasf}) we find that for all these solutions the relation
$\epsilon=0$ holds. Looking at the expressions of the free energy
and of the saddle-point equations (\ref{SPeq}) we see that $\epsilon=0$ implies that these solutions solve the unconstrained two-system problem too. Now, it is well known that the $P(q)$ of the
single-system problem can be reconstructed from the matrix $Q$
used to evaluate the free energy \cite{MPV}, one finds the
following relations
\begin{equation}
q^{(k)} = \int q^{k}P(q)dq = \lim_{n\rightarrow
0}\overline{Q_{\alpha\beta}^{k}}\label{PQ}
\end{equation}
where the bar denotes average over all the solutions of the
saddle-point equation . Indeed when replica symmetry is broken
there is always more than one solution; given one of them others
can be obtained by permutations of the replica indices.
Following the same steps one can show that the function
$P(q_{T1T2})$ can be obtained from the order parameter
 $\hat{Q}=\left(\begin{array}{cc}
 Q_{1} & P \\
 P^{t} & Q_{2}
 \end{array}\right)$
of the {\em unconstrained} two-system problem through the
relations
\begin{equation}
\label{p12}
q_{T1T2}^{(k)} = \int q_{T1T2}^{k}P(q_{T1T2})dq_{T1T2} =
\lim_{n\rightarrow 0}\overline{P_{\alpha\beta}^{k}}\label{PQ12}
\end{equation}
 Again the bar means average
over all the solutions of the saddle point equations; we have of
course the solution $P=0,\ Q_{1}=Q_{1free},\ Q_{2}=Q_{2free}$, but
contrary to what sometimes stated nothing forbids the existence of
solutions with a nonzero $P$. By looking at (\ref{p12}) we see
that we may not know all the solutions or how to average over
them but what we do know is that the existence of solutions with $P\neq 0$ implies a nontrivial $P(q_{T1T2})$. 

As stated above, solutions
of the constrained two-system model with the same free energy of
the unconstrained one are solutions of the latter as well, so we
can safely say that if these solutions exist there is no chaos in
temperature, while, if they don't, pure states at different temperatures
are completely uncorrelated.

The fact that non-chaotic solutions of the constrained two-system problem are also solutions of the unconstrained problem is very important.
As we saw, by this argument we can safely claim that if these solutions exist the $P(q_{T1T2})$ is non trivial.
Instead it will be dangerous to infer a non trivial $P(q_{T1T2})$ only from the fact that the free energy per spin in the constrained case is equal to the sum of the free energy of the two uncoupled systems. We shall describe a situation where this recipe could lead to wrong results.

The the spin-glass pure states have all the same free energy per spin but their relative weights, due to the correction of order $1/N$, are very different and only few states are relevant for physical quantities like the $P(q)$. 

Now let us imagine a situation of weak chaoticity where the two complete sets of states at different temperatures are strongly correlated but the corrections of order $1/N$ are completely reshuffled: the few relevant states at $T_2$ are located in a different portion of the phase space from that of the relevant states at $T_1$. Therefore in this situation we must have a trivial $P(q_{T1T2})$, i.e. a delta function centered in zero. At the same time we will also have (non-relevant) states with a non-zero overlap between them. A non-zero constraint will select those couples of non-relevant states yielding the same free energy per spin of the unconstrained case, inducing one wrongly to think that the support of the $P(q_{T1T2})$ is non-zero, i.e. that there is no chaos in temperature, while we assumed at the beginning that chaos in temperature, although weak, is present.

The fact that there is a direct connection between the constrained and unconstrained cases ensures that the weak chaoticity picture described above is not possible and that by studying the constrained problem we always obtain sound physical information on the function $P(q_{T1T2	})$ of two uncoupled systems.

Let us discuss the results reported in \cite{BY}; rephrased in the context we have discussed, the method consists in studying the unconstrained problem expanding
the free-energy functional (\ref{F}) as follows
\begin{equation}
F(\hat{Q})=F^{(0)}(\hat{Q})+\Delta T F^{(1)}(\hat{Q})+\Delta T^2
F^{(2)}(\hat{Q})+\ldots\label{dtexpansion}
\end{equation}
Where $F^{(0)}$ is the functional at equal temperatures and
$\Delta T$ the temperature difference assumed to be small.

The unperturbed equal-temperature case is readily solved: the
Parisi solution in its standard single-system form corresponds to
$Q_1=Q_{standard},\ Q_2=Q_{standard},\ P=q(0)=0$. The
symmetry between the two systems definitely implies that solutions with a
non-zero $P$ can be obtained from the standard one by
proper permutations of the replica indeces. Now, the perturbing
terms in (\ref{dtexpansion}) break the symmetry between the two
systems and one has to check whether they remove the degeneracy
in free energy of the solutions. 

The free energy of each solution
is evaluated in powers of $\Delta T$ substituting
in (\ref{dtexpansion}) the zero-order equal-temperature
solutions; by this procedure a free-energy difference to
second order in $\Delta T$ is found. However, {\em to second order
one must consider the contribution to $F$ that belongs to the
splitting of the zero-order solution}. Explicitly one must add
the term
\begin{equation}
\Delta T^2{\partial F^{(1)} \over \partial
\hat{Q}_{ab}}({\mathcal{D}} ^2 F^{(0)} )_{ab,cd}^{-1}{\partial
F^{(1)} \over
\partial \hat{Q}_{ab}}\left|_{\hat{Q}=\hat{Q}^{(0)}}\right.\label{expterm}
\end{equation}
This term too is not permutationally invariant; therefore it might cancel
the one belonging to the expansion of $F$.
While $F^{(1)}$ has a simple expression evaluating exactly (\ref{expterm}) it's impossible to the present knowledge. 

However, the whole procedure seems unreliable for more
general reasons. For instance, if we try to apply it to the
standard single-system case we face the problem of the
non-analytical temperature dependence of the Parisi solution $q(x)$
caused by the presence of the plateau. If we think in terms of the
hierarchical replica-symmetry breaking (RSB) scheme it can be realized that by trying to construct a solution starting from a
different-temperature one we obtain a solution with the same
break point i.e. a wrong one.

\section{Non-Chaotic Solutions}

Near the critical temperature the order parameter is expected to
be small so that one can expand the SP equations (8) in
powers of $\hat{Q}$ and then obtain approximate solutions in
powers of the reduced temperature $\tau=T_{c}-T$.

By means of standard manipulations and a proper
temperature-dependent rescaling of the order parameter
($\beta_1^2Q_1\rightarrow Q_1,\beta_1^2Q_2\rightarrow Q_2,\beta_1\beta_2P\rightarrow P$) we obtain the free energy
up to fifth order
\begin{eqnarray}
F(\hat{Q})=-\lim_{n\rightarrow 0}\frac{1}{2}\{\tau_{1}\Tr
Q_{1}^{2}+\tau_{2}\Tr Q_{2}^{2}+2\tau_{12}\Tr
P^{2}+\frac{\omega}{3}\Tr \hat{Q}^{3} \nonumber
\\
+\frac{u}{6}\sum_{ab}\hat{Q}_{ab}^{4}+\frac{v}{4}\Tr
\hat{Q}^{4}-\frac{y}{2}\sum_{abc}\hat{Q}_{ab}^{2}\hat{Q}_{bc}^{2}+\frac{z}{5}\Tr
\hat{Q}^{5} \nonumber
\\
\label{Fexp}
-s\sum_{ab}\hat{Q}^{2}_{ab}(\hat{Q}^{3})_{aa}+\frac{2}{3}t\sum_{ab}\hat{Q}^{3}_{ab}(\hat{Q}^{2})_{ab}+\frac{n}{2}(p_{d}-q_{c})^{2}\}
\end{eqnarray}
Where
\begin{eqnarray}
\tau_{1}=\frac{1-T_{1}^{2}}{2}\ \ \ \ \ \ \
\tau_{2}=\frac{1-T_{2}^{2}}{2}\ \ \ \ \ \ \
\tau_{12}=\frac{1-T_{1}T_{2}}{2}
\end{eqnarray}

In the SK model we have $\omega =u=v=y=z=s=t=1$. The only term
that explicitly depends on the constraint is
$(p_{d}-q_{c})^{2}=\epsilon^{2}$, so that a general strategy
often employed is to fix the diagonal term in $P$ and to solve
the saddle-point equations belonging to the remaining components
in $\hat{Q}$. By this procedure we obtain a solution
corresponding to the constraint
\begin{equation}
q_{c}=p_{d}-\frac{\partial F_{free}(\hat{Q})}{\partial p_{d}}
\end{equation}
We skip the expressions of the SP equations expanded to fifth order in $Q_1,Q_2$ and $P$. To a lower order they are reported in equations (\ref{SPTOT}) and (\ref{SPQ1}) . 

A first attempt to solve the equations was to set all the
components of $P$ as constant, this ansatz has a positive
free-energy cost, i.e. chaos, {\em independently of the
temperature difference}; however we know that non-chaotic
solutions exist for equal temperatures \cite{FPV1}, for this
reason we have tried an ansatz for $\hat{Q}$ that could reproduce
these solutions in the limit of equal temperatures. 
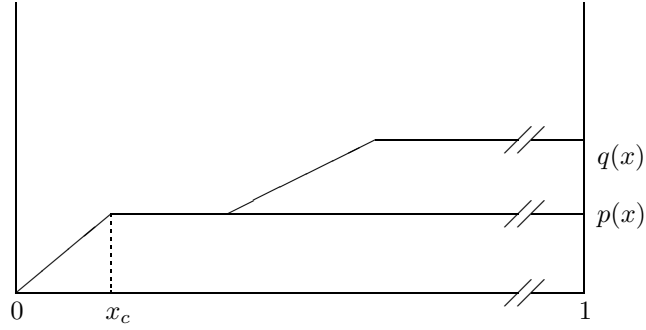
\begin{figure}[bt]
\begin{center}
\begin{picture}(280,130)(-27,-12)
\put(0,0){\line(0,1){110}} \put(0,0){\line(1,0){190}}

\put(0,0){\line(6,5){36}}

\multiput(36,30)(0,-3){10}{\line(0,-1){1}}
\put(36,0){\line(0,1){1}} \put(34,-10){$x_c$}

\put(36,30){\line(1,0){154}}

\put(80,30){\line(2,1){56}}

\put(136,58){\line(1,0){54}}

\put(185,53){\line(1,1){10}}\put(185,25){\line(1,1){10}}\put(185,-5){\line(1,1){10}}

\put(190,53){\line(1,1){10}}\put(190,25){\line(1,1){10}}\put(190,-5){\line(1,1){10}}

\put(195,58){\line(1,0){20}}\put(195,30){\line(1,0){20}}\put(195,0){\line(1,0){20}}

\put(215,0){\line(0,1){110}}

\put(-2,-10){0}\put(213,-10){1}

\put(220,49){$q(x)$}\put(220,27){$p(x)$}
\end{picture}
\caption{a qualitative sketch of the solutions in the isothermal case: $q_1(x)=q_2(x)=q(x)$ for all $x$; $p(x)=q(x)=q_{free}(2x)$ for $x<x_c$, for $x>x_c$ $p(x)$ remains constant and equal to $q_c$ while $q(x)$ after an intermediate plateau is joined continuously to $q_{free}(x)$; $q_c$ can take value between
zero and $q_{EA}$. The diagonal value of $P$ is equal to the constraint, i.e. $p_d=q_c$} \label{figure1}
\end{center}
\end{figure}

In the case
of equal temperatures the solutions of the constrained problem are represented in figure (\ref{figure1}); they 
display full Replica-Symmetry Breaking (RSB) in the matrix $P$
and have a rather simple structure 
\begin{eqnarray}
q_{2}(x)=q_{1}(x);p_{d}=q_{c} & \textrm{for\ all\ }x \nonumber
\\
 q_{1}(x)=p(x)=q_{free}(2x) & 0\leq x\leq \frac{1}{2}x_{free}(p_{d})\nonumber
 \\
 q_{1}(x)=p(x)=p_{d} &
 \frac{1}{2}x_{free}(p_{d})\leq x\leq x_{free}(p_{d})\nonumber
 \\
\label{eqts}
 q_{1}(x)=q_{free}(x);p(x)=p_{d}& x_{free}(p_{d})\leq x\leq 1
\end{eqnarray}
where $q_{free}(x)$ is the free Parisi solution, and $x_{free}(q)$
is its inverse. The diagonal terms in $P$ are equal to the constraint, i.e. $q_c=p_d$. These solutions exist for any value of $q_{c}$ in
the support of the function $P(q)$ and have exactly the same free
energy of the free case; actually it can be shown that they are
particular permutations of the free single-system solution
\cite{inprep}.

We want to remark that the structure of these solutions is {\em
intrinsically}\ non-chaotic due to the equality $p(x)=p_{d}$ for $x>x_c$.
Indeed the SP equations read
\begin{eqnarray}
Q_{1\alpha\beta}=\langle
S^{1}_{\alpha}S^{1}_{\beta}\rangle&Q_{2\alpha\beta}=\langle
S^{2}_{\alpha}S^{2}_{\beta}\rangle \nonumber
\\
P_{\alpha\beta}=\langle S^{1}_{\alpha}S^{2}_{\beta}\rangle\
+\frac{1}{\beta_{1}\beta_{2}}\/\epsilon_{\alpha}\delta_{\alpha\beta}&\epsilon_{\alpha}=\beta_1\beta_2(P_{\alpha\alpha}-q_{c})
\end{eqnarray}
Now if we have $p(1)=p_{d}$ we can make the permutation
$S^{1}_{1}\leftrightarrow S^{1}_{2}$, which is {\em not} a natural
permutation of the Hamiltonian, leaving the matrix $\hat{Q}$
unchanged. Since $\epsilon$ appears only in the equation for the diagonal term $P_{\alpha\alpha}$ (i.e. $p_d$ within the Parisi ansatz), this implies
\begin{equation}
p(1)=\langle S^{1}_{1}S^{2}_{2}\rangle=\langle
S^{1}_{2}S^{2}_{2}\rangle=p_{d}\rightarrow\ \epsilon=0\label{epseq}
\end{equation}
So we see that $\frac{\partial{F(q_{c})}}{\partial{q_{c}}}=0$ is
true for all these solutions and noticing that they approach
continuously the free solutions in the limit of zero constraint we can
safely claim that all these solutions have the same free-energy of
the free problem.

As stated above, we have looked for solutions approaching
continuously (\ref{eqts}) in the limit of equal temperatures, so
we made some variational attempts with finite RSB for the matrix
$P$. Two facts emerged from this analysis: i) if we allow
$p(1)\neq p_{d}$ the variational solution is always attracted by
a solution with a positive free-energy cost {\em independently}
of the temperature difference, so we discard it; ii) if we force
the variational trial function to have $p(1)=p_{d}$ we always
obtain a negative free-energy cost. 

This last result is absurd
from a physical point of view since by imposing the constraint we
are reducing the configuration space, but in a variational
computation it may occur if our trial function doesn't approach
enough the true maximum (one of the subtleties of the replica
trick is that one has to maximize and not minimize the free
energy). The fact that even very complex variational functions
showed negative free energy cost made us suspect that the maximum
had zero free-energy cost, so we turned to directly solving the
saddle-point equation near the critical temperature.
\begin{figure}[btp]
\begin{center}
\begin{picture}(280,130)(-27,-12)
\put(0,0){\line(0,1){110}} \put(0,0){\line(1,0){190}}

\put(0,0){\line(1,1){36}} \put(0,0){\line(6,5){36}}
\put(0,0){\line(3,2){36}}

\multiput(36,36)(0,-3){12}{\line(0,-1){1}}
\put(36,0){\line(0,1){1}} \put(34,-10){$x_c$}

\put(36,36){\line(1,0){39}} \put(36,30){\line(1,0){154}}
\put(36,24){\line(1,0){44}}

\put(75,36){\line(2,1){80}} \put(80,24){\line(2,1){56}}

\put(155,76){\line(1,0){35}} \put(136,52){\line(1,0){54}}

\put(185,71){\line(1,1){10}}
\put(185,47){\line(1,1){10}}\put(185,25){\line(1,1){10}}\put(185,-5){\line(1,1){10}}

\put(190,71){\line(1,1){10}}
\put(190,47){\line(1,1){10}}\put(190,25){\line(1,1){10}}\put(190,-5){\line(1,1){10}}

\put(195,76){\line(1,0){20}}
\put(195,52){\line(1,0){20}}\put(195,30){\line(1,0){20}}\put(195,0){\line(1,0){20}}

\put(215,0){\line(0,1){110}}

\put(-2,-10){0}\put(213,-10){1}

\put(220,73){$q_2(x)$}\put(220,49){$q_1(x)$}\put(220,27){$p(x)$}
\end{picture}
\caption{a qualitative sketch of the solutions, in the small-$x$
region they are all different until the point $x_c$ where
$p(x)=p_d=q_c$, for $x>x_c$ $p(x)$ is constant and equal to $q_c$, while $q_1(x)$ and
$q_2(x)$ after an intermediate plateau are joined continuously to
the corresponding free function; $q_c$ can take values between
zero and some $q_{c max}$ at which the two plateaus of the
function at the higher temperature merge. At zero order the slope
of the functions is 1 in the first region and 1/2 in the
intermediate ones.} \label{figure}
\end{center}
\end{figure}
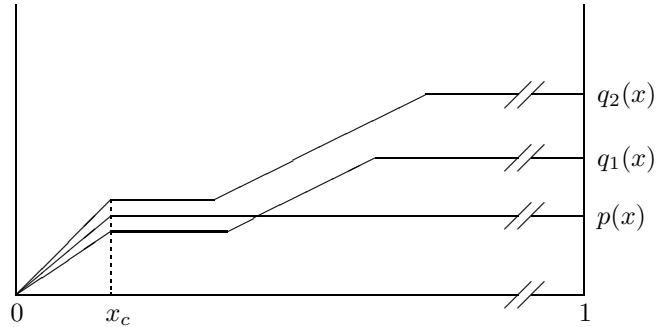

Having in mind the equal temperature case, we looked at solutions
with the structure depicted in figure (\ref{figure}): in the
small-$x$ region the three functions $q_{1}(x),q_{2}(x)$ and
$p(x)$ are all different till they reach the point $x_{c}$ where
$p(x)=p_{d}=q_{c}$; then for $x$ greater than $x_{c}$ $p(x)$
remains constant while $q_{1}(x)$ and $q_{2}(x)$ after an
intermediate plateau are connected continuously to the
corresponding free Parisi solutions. These solutions are thought to exist for
values of the constraint from zero to a maximum value where the
two plateaus of the function at the higher temperature merge ($T_1>T_2$ in figure (\ref{figure})). 

We postpone to section 5 the discussion of a consistency problem concerning such a structure for the solutions.

In the isothermal case the functions in the small-$x$ region do not
depend on the value of $q_c$, which only acts as a knife that fixes
the position and length of the intermediate plateaus; to the
order we compute we cannot say if this is true even for the
two-temperature problem, but we believe that it is.
It is important to notice that these solutions, provided they actually exist, intrinsically display a zero free-energy cost due to the equality $p(1)=p_d$ as in the isothermal case 
(see equation (\ref{epseq})).

We checked that to the fifth order in the expansion in the order parameter it is possible to find non-chaotic solutions with the structure described above. In the following we sketch the essential features of the calculation.

The expression for the free-energy
functional truncated to the fifth order allows us to compute the
functions to the second order in the regions before the starting
point of the large plateau because  these regions already span an interval of
order one in the reduced temperatures. 
Instead, the value of the plateau can be evaluated to the
third order as it spans a region of order zero in the reduced temperature.

 The saddle-point equations can be solved in terms of the Parisi functions
$q_{1}(x),q_{2}(x)$ and $p\/(x)$ expanded in powers of $x$; one can then use the
terms proportional to $x^{3}$ and higher to determine the
functions in the small-$x$ region and the terms proportional to
$x$ to determine the value of the plateau. 
We determined the
functions in the small-$x$ region and then checked such 
functions joined with the free Parisi solutions to satisfy the equations. 

To the fifth order in $\hat{Q}$ the SP equations display terms
proportional to $x$ and to $x^{3}$, plus constant terms due to the
presence of the intermediate plateau for $x>x_{c}$.  Having
determined the functions in the small $x$ region using the
$x^{3}$ terms we had to check nine coefficients to be zero: three
proportional to $x$, one for each of the three equations for
$q_{1}(x),q_{2}(x)$ and $p\/(x)$ in the small-$x$ region; two
proportional to $x$ in the regions of the intermediate $x$ from
the two equations for $q_{1}(x),q_{2}(x)$; four belonging
to the same regions corresponding to the $x^{3}$ and constant terms.

After a tedious but straightforward computation all these
coefficients turned out to be zero. However, while the coefficients of $x,x^3,\ldots$ in the
equations for $q_{1}(x)$ and $q_{2}(x)$ are null independently of
the value of $\omega,u,v,y$\ldots the corresponding coefficients in the
equation for $p\/(x)$ only cancel provided the following relationships
hold
\begin{equation} \label{cancrel}
1-\frac{v}{\omega^{2}}=0\ \ \ \ \ \ \ \ 1-\frac{3
v^{2}}{\omega^{4}}+\frac{2 z}{\omega^{3}}-\frac{2 v
y}{\omega^{4}}+\frac{2 s}{\omega^{3}}=0
\end{equation}
The constant terms in these relationships are the coefficients of
$(\tau_{1}-\tau_{2})^{2}$ and
$(\tau_{1}+\tau_{2})(\tau_{1}-\tau_{2})^{2}$ in the expansion of
$4\tau_{12}$. 

The free-energy difference 
has been evaluated either directly on the maximum and both
thermodynamically integrating the energies (that have a much
simpler expression in the order parameter) with respect to the
temperatures. It turned out to be proportional to the left-hand sides of (\ref{cancrel}), so that it is zero when relations (\ref{cancrel}) hold. This is consistent with the previous statement that the
structure of the solutions is intrinsically non-chaotic; indeed, it implies that we cannot choose the coefficients
$\omega,u,v,y$\ldots to have a positive free-energy cost and at
the same time to satisfy the saddle point equations. 

The expressions of $q_{1}(x),q_{2}(x)$ and $p\/(x)$ in terms of
$\omega,u,v,y$\ldots are reported in the appendix. 

\section{Exact Results: the Constraint-Independent Cancellations}

In this section we prove that a particular set of cancellations in the SP equation of $p(x)$ is verified at all orders in the SK model. Expanding the equation in powers of $x$, we must find that the coefficient of each power equals zero. We concentrate on the coefficient of $x$, and in particular on those terms in the coefficient of $x$ that do not depend of the value of the constraint $q_c$, but only on the temperatures of the two systems. These terms must sum up to zero. It will be proven that these constraint-independent cancellations hold at any order in the SK model.
We will see that these cancellations are the origin of the relationship (\ref{cancrel}); this means that going to higher order we will encounter relations between the coefficient like (\ref{cancrel}) such that they are verified by the corresponding coefficients of the SK model.

One should not forget that to verify the SP equations others cancellations should be checked at all order, while they have been checked only to lowest orders in the computation reported above; however it is interesting to notice that at the order we computed we found that these cancellations hold independently of the parameters $\omega,u,v,y$\ldots so one may conjecture that at higher orders they can always be fullfilled by opportunously tuning the functions in the small-$x$ region. According to this conjecture for any given spin-glass model the only relevant condition for the existence of the non-chaotic solutions is that the constraint-independent cancellations we consider here hold at all orders.

The free energy (\ref{Fexp}) can be written as the sum of the
free-energies of the free problem plus a  term proportional to
$P$ which reads
\begin{eqnarray}
F(P)=-\lim_{n\rightarrow 0}\frac{1}{2n}\{ \Tr (AP^{2})
+\frac{u}{3}\sum_{ab}P_{ab}^{4}+\frac{v}{2}\Tr
P^{4}-y\sum_{abc}P_{ab}^{2}P_{bc}^{2}\nonumber \\+z \Tr
P^{4}(Q_{1}+Q_{2})-\frac{3s}{n}\Tr P^{2}(Q_{1}+Q_{2})\Tr
P^{2}+\frac{4}{3}t\sum_{ab}P_{ab}^{3}(PQ_{1}+PQ_{2})_{ab}\}\label{FP}
\end{eqnarray}
\begin{equation}\label{A}
\begin{array}{rl}
A_{ab}&=(2\tau_{12}-{y\over n}(\Tr Q_{1}^{2}+\Tr
Q_{2}^{2})-{s\over n}(\Tr Q_{1}^{3}+\Tr
Q_{2}^{3}))\delta_{ab}+\omega(Q_{1}+Q_{2})_{ab}
\\
&+v(Q_{1}^{2}+Q_{2}^{2}+Q_{1}Q_{2})_{ab}+z(Q_{1}^{3}+Q_{2}^{3}+Q_{1}Q_{2}^{2}+Q_{2}Q_{1}^{2})_{ab}
\\
&-{s\over n}(\Tr Q_{1}^{2})(2Q_{1}+Q_{2})_{ab}-{s\over n}(\Tr
Q_{2}^{2})(2Q_{2}+Q_{1})_{ab}+{2\over 3}t(Q_{1ab}^{3}+Q_{2ab}^{3})
\end{array}
\end{equation}
The equation one obtains differentiating with respect to $P$ is
\begin{equation}\label{eqP}
-\frac{\partial F(P)}{\partial P_{ab}}=(AP)_{ab}+{2\over 3}u
P^{3}_{ab}+v(P^{3})_{ab}-yP_{ab}((P^{2})_{aa}+(P^{2})_{bb})+{1\over
2}\epsilon\ \delta_{ab}=0
\end{equation}
In the last equation we neglected the fourth order terms which
contain powers of $P$ greater than one as in reality their order
is higher. We are assuming that $q_{c}$ is of the order of
magnitude of the self overlap in the free case i.e. of the order
of $\tau_{1}$ and $\tau_{2}$. The structure of the matrix $P$ (explicitely the equality $p_d=p(1)$)
implies that even if $P$ is of order one its powers $P_{n}$ are of
order $2n-1$ and not of order $n$. Being $(p^{(n)}_{d},p^{(n)}(x))$ the Parisi
function associated to $P^{n}$, we know that the quantity
$p^{(n)}_{d}-\bar{p}^{(n)}$ equals $(p_{d}-\bar{p})^{n}$ i.e. it's
proportional to $q_c^{2n}$. Now $p^{(n)}(x)$ is different from
$p_{d}^{(n)}$ only on a region of order one and we see immediately
that $P_{n}$ must be of order $2n-1$ in $q_{c}$.

Expressing (\ref{eqP}) in $q_1 (x),q_2(x),p(x)$ and expanding in
powers of $x$ we have terms proportional to $x$ and $x^{3}$ in the
small-$x$ region. We determined the function $p(x)$ using the
$x^3$ terms and thus we must check the $x$ ones to be zero, these are
generated by
 $(AP)_{ab}$ and $P_{ab}((P^{2})_{aa}+(P^{2})_{bb})$.
 Then the equation we have to check is
\begin{equation}
\label{pequa}
(a_{d}-\bar{a})p(x)+(p_{d}-\bar{p})a(x)+2(p_{d}^{2}-\bar{p^{2}})p(x)=0
\end{equation}
where $(a_{d},a(x))$ is the Parisi function associated to the
matrix $A$ defined in (\ref{A}). This equation must really hold
only for the terms proportional to $x$ and we may more
correctly write it substituting $q(x)$ and $p(x)$ with their
derivative in zero eliminating the $x$-dependence.

 The l.h.s. of
(\ref{pequa}) displays terms explicitly proportional to powers of
$q_c$ and others that depend {\em only} on $\tau_{1}$ and
$\tau_{2}$; the former terms cancel irrespective of the values of
$\omega,u,v,y\ldots$ while the latter sum up to zero provided that
the relations (\ref{cancrel}) hold.

 Since at zero order we have $p'(0)=1$ it is easily seen that
these constraint-independent terms belong only to
$(a_{d}-\overline{a})$. It is quite natural that their
cancellation depends crucially on the structure of the model
(i.e. on the coefficients $\omega ,u,v,y\ldots$) because they
cannot be controlled by properly tuning the solutions in the
small-$x$ region. For completeness we report the
$q_c$-independent terms belonging to the quantity $a_{d}-\bar{a}$
evaluated directly from (\ref{A}) which is the origin of the
relationships (\ref{cancrel})
\begin{equation}
a_{d}-\bar{a}={1\over
2}\left(1-\frac{v}{\omega^{2}}\right)(\tau_{1}-\tau_{2})^{2}+{1\over
2}\left(1-\frac{3 v^{2}}{\omega^{4}}+\frac{2
z}{\omega^{3}}-\frac{2 v y}{\omega^{4}}+\frac{2
s}{\omega^{3}}\right)(\tau_{1}+\tau_{2})(\tau_{1}-\tau_{2})^{2}
\end{equation}
 We will prove that the constraint-independent terms in $a_{d}-\bar{a}$ are null at all orders in
 the SK model.
 
According to the previous discussion 
the matrix $A$ is defined as the sum of all the $P$-independent matrices $X$ that appears in the free-energy functional in the form Tr$XP^{2}$ (actually the expression (\ref{A}) is truncated to third order in the order parameter, because it is derived from (\ref{Fexp}) which is in turn the free-energy functional truncated to the fifth order). Therefore $A$ it's related to the second derivative of $F$ with respect to
$P$. By direct inspection one verifies that
$A_{am}\delta_{bn}$ is equal to ${\partial^{2} F}/{\partial
P_{ab}\partial P_{mn}}$ plus terms that depend explicitly on $P$.
We are interested in the $q_c$-independent part of $A$ so
we can consider the limit $q_c \rightarrow 0$ where they
become equal. Therefore we have to compute ${\partial^{2}
F}/{\partial P_{ab}\partial P_{mn}}$ in the limit of zero
constraint (i.e. $P\rightarrow 0$). The
solutions reduce continuously to the free ones in this limit so we can safely compute it for $P=0$. 

Since ${\partial^{2}
F}/{\partial P_{ab}\partial P_{mn}}$ is a component of the Hessian
of $F$, it is the main ingredient to calculate the
spatial correlation functions between states at different
temperatures of a finite-dimensional spin-glass in a Gaussian approximation
 around mean-field theory. In this context it has been previously computed by Kondor \cite{Kondor1}; therefore
{\em the particular cancellations we are dealing with are exactly
verified at all orders for much the same reason that at zero-loop
order the overlap of the spin correlations at two different
temperatures is infinitely long ranged, whatever the difference
between them.} 

One can easily convince himself that, when
$P=0$, $(a_{d}-\bar{a})$ is proportional to the minimum eigenvalue
of ${\partial^{2} F}/{\partial P_{ab}\partial P_{mn}}$ i.e. to the
inverse of the correlation length. Therefore the computation we
shall sketch below is a completely equivalent rephrasing of
Kondor's computation.

Let us proceed to evaluate $(a_{d}-\bar{a})$; following Kondor, we
start noticing that the Hessian is related to the four-point
connected correlation functions
\begin{equation}
nT_{1}T_{2}A_{am}\delta_{bn}=nT_{1}T_{2}\frac{\partial^{2}
F}{\partial P_{ab}\partial
P_{mn}}=\delta_{ab}\delta_{mn}-\frac{1}{T_{1}T_{2}}\left( \langle
S_{1}^{a}S_{2}^{b}S_{1}^{m}S_{2}^{n} \rangle-\langle
S_{1}^{a}S_{2}^{b} \rangle\langle S_{1}^{m}S_{2}^{n}\rangle\right)
\label{kondrel}
\end{equation}
The first identity holds because $P$ is zero; exploiting again
this relation we can evaluate the r.h.s. of (\ref{kondrel})
\begin{eqnarray}
nT_{1}T_{2}A_{am}\delta_{bn}=\delta_{ab}\delta_{mn}-\frac{1}{T_{1}T_{2}}(\delta_{am}\delta_{bn}+\delta_{am}(1-\delta_{bn})Q_{2bn}
\nonumber
\\
+\delta_{bn}(1-\delta_{am})Q_{1am}+(1-\delta_{am})(1-\delta_{bn})Q_{2bn}Q_{1am})
\end{eqnarray}
\begin{equation}
A_{ab}=\sum_{mn}(A_{am}\delta_{bn})\delta_{mn}=\frac{1}{T_{1}^{2}T_{2}^{2}}(T_{1}T_{2}-1-Q_{1ab}-Q_{2ab}-(Q_{1}Q_{2})_{ab})
\end{equation}
Now, given a generic ultrametric matrix $A$, the quantity
$a_{d}-\bar{a}$ is the eigenvalue corresponding to the
eigenvector with constant coordinates so we have
\begin{equation}
a_{d}-\bar{a}=\frac{1}{T_{1}^{2}T_{2}^{2}}(T_{1}T_{2}-1+\bar{q}_{1}+\bar{q}_{2}-(\bar{q}_{1})(\bar{q}_{2}))=0
\end{equation}
Where in the last identity we used the exact relation \cite{Sommers}
 $\bar{q}_{Parisi}=1-T$. Note that this relation holds because we have assumed that in the large-$x$ region the functions $q_1(x)$ and $q_2(x)$ are exactly equal to the standard Parisi solutions at the corresponding temperatures.

\section{Structural Consistency}

In this section we discuss a problem of consistency related to the structure we propose for the solutions. The essential features of the problem have been presented in the introduction. It can be view as a necessary, but non sufficient, condition to solve the SP equations. We will show that this necessary condition is fullfilled at all orders by the solutions we propose. As said in the introduction this result is completely independent on the result of the previous section.

In the following we will use the SP equations expanded to the lowest order sufficient to clarify the nature of the problem but the results we will obtain are valid at all orders.
The saddle-point equations for the single system problem read
\begin{equation}
2\tau_1Q_{ab}+(Q^2)_{ab}+{2\over 3}Q_{ab}^3+(Q^3)_{ab}-Q_{ab}((Q^2)_{aa}+(Q^2)_{bb})+\ldots=0\label{SPSS}
\end{equation}
the l.h.s. is a sum of all the possible covariants of $Q$. A covariant is a two-index object built in a permutational covariant way from the matrix $Q$; given any permutation $\pi$ between the replica indeces the mathematical definition of a covariant $M$ is 
\begin{equation}
M(\pi Q)=\pi M(Q)\label{defcov}
\end{equation}
If the matrix $Q$ is a hierarchical matrix, all its covariants are hierarchical matrices with the same structure of Q, i.e. with the same set of block indexes $m$'s; for instance, if the Parisi function $q(x)$ associated to $Q$ has a plateau for $x$ greater then some $x_1$ the Parisi functions associated to any of its covariants will display a plateau in the same region. 
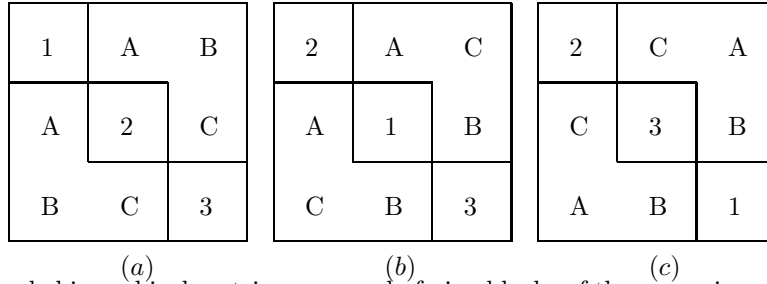
\begin{figure}[btp]
\begin{center}
\begin{picture}(290,90)(0,0)

\put(0,0){\dashbox{90}(90,90)}
\multiput(0,60)(30,-30){3}{\dashbox{30}(30,30)}

\put(100,0){\dashbox{90}(90,90)}
\multiput(100,60)(30,-30){3}{\dashbox{30}(30,30)}

\put(200,0){\dashbox{90}(90,90)}
\multiput(200,60)(30,-30){3}{\dashbox{30}(30,30)}

\put(42,-13){$(a)$}
\put(12,10){B}
\put(12,40){A}
\put(12,70){1}
\put(42,10){C}
\put(42,40){2}
\put(42,70){A}
\put(72,10){3}
\put(72,40){C}
\put(72,70){B}

\put(142,-13){$(b)$}

\put(112,10){C}
\put(112,40){A}
\put(112,70){2}
\put(142,10){B}
\put(142,40){1}
\put(142,70){A}
\put(172,10){3}
\put(172,40){B}
\put(172,70){C}

\put(242,-13){$(c)$}

\put(212,10){A}
\put(212,40){C}
\put(212,70){2}
\put(242,10){B}
\put(242,40){3}
\put(242,70){C}
\put(272,10){1}
\put(272,40){B}
\put(272,70){A}

\end{picture}
\caption{$(a)$ A simple hierarchical matrix composed of nine blocks of the same size with 1=2=3=$q_D$ and A=B=C=$q_{OD}$. The matrix is invariant under block permutations like those represented in $(b,c)$, they can be used to prove separability.} \label{figure3}
\end{center}
\end{figure}

This property of the Parisi matrices can be understood by looking at figure (\ref{figure3}): we have a simple hierarchical matrix $Q$ composed of nine blocks of the same size; the components of a single block are all equals, the three blocks on the diagonal are equal to $q_D$ and  the six blocks off the diagonal are equal to $q_{OD}$. 

In figure (\ref{figure3}) two blocks permutations are also represented which leave the matrix $Q$ unchanged; these permutations can be used to prove that any covariant of $Q$ has its same structure. For instance, if we consider the first permutation and apply to it the definition of covariant (\ref{defcov}) together with the fact that this permutation verifies $\pi (Q)=Q$, we obtain that $\pi M=M$; Since this permutation exchanges blocks $1$ and $2$, the blocks corresponding to $1$ and $2$ in $M$ are equal, as the blocks corresponding to $B$ and $C$; using other permutations of this kind we can prove 
that any covariant $M$ is a hierarchical matrix with the same structure of $Q$.

This property goes under the name of separability or overlap equivalence \cite{SEP} because of its physical meaning; it is not clear if it's a condition to solve the saddle-point equation, but, together with stochastic stability, it's likely to be the origin of ultrametricity.

The SP equation for two systems at different temperatures read
\begin{equation}
2\left(\begin{array}{cc}
\tau _{1} Q_{1} & \tau_{12}P
\\
\tau_{12}P & \tau _{2} Q_{2}
\end{array}\right)_{ab}
+(\hat{Q}^2)_{ab}+{2\over 3}\hat{Q}_{ab}^3+(\hat{Q}^3)_{ab}+\ldots=0;\end{equation}
Where the indeces $(a,b)$ range from $1$ to $2n$. All the terms -excluding the first- in the previous equation are identical to those of the SP equation for a single system (\ref{SPSS}), provided one substitute the covariants of $Q$ with the corresponding covariants of $\hat{Q}$. 
Expressing the covariants of $\hat{Q}$ in terms of $Q_1,Q_2$ and $P$ the previous equation reads
\begin{displaymath}
2\left(\begin{array}{cc}
\tau _{1} Q_{1\ ab} & \tau_{12}P_{ab}
\\
\tau_{12}P_{ab} & \tau _{2} Q_{2\ ab}
\end{array}\right)+\left(\begin{array}{cc}
(Q_{1}^2+P^2)_{ab} & (P(Q_1+Q_2))_{ab}
\\
(P(Q_1+Q_2))_{ab} & (Q_2^2)_{ab}
\end{array}\right)+{2\over 3}\left(\begin{array}{cc}
Q_{1\ ab}^3 & P_{ab}^3
\\
P_{ab}^3 & Q_{2\ ab}^3
\end{array}\right)
+
\end{displaymath}
\begin{equation}
+\left(\begin{array}{cc}
(Q_1^3+P^2(2Q_1+Q_2))_{ab} & (P^3+P(Q_1^2+Q_2^2+Q_1Q_2))_{ab}
\\
(P^3+P(Q_1^2+Q_2^2+Q_1Q_2))_{ab}& (Q_2^3+P^2(2Q_2+Q_1))_{ab}\end{array}\right)+\ldots=0
\label{SPTOT}
\end{equation}
Where the indeces $(a,b)$ range from $1$ to $n$. According to the previous equations we have that the equation for $Q_1$ reads
\begin{eqnarray}
2\tau_1Q_{1ab}+(Q_1^2)_{ab}+P^2_{ab}+{2\over 3}Q_{1ab}^3+(Q_1^3)_{ab}+((2Q_1+Q_2)P^2)_{ab}+\ldots=0\label{SPQ1}
\end{eqnarray}
Notice that $P$ always
appears in even power in the invariants of the free-energy, this is connected with the fact that  each mute index of $\hat{Q}$ must appears an even number of times in zero
magnetic field.

The consistency problem we want to address is that the previuos equations do not admit a solution for any parametrization of the matrices $Q_1,Q_2$ and $P$.

Let us start considering the simplest case where $Q_1,Q_2$ and $P$ are 1RSB Parisi matrices with different breaking points. If we want to solve (\ref{SPQ1}) to second order we must consider its first three terms. The first two are 1RSB matrices with breaking point $x_{Q_1}$, while the second is a 1RSB matrix with breaking point $x_{P}$. If they are to sum up to zero the only two possibility are: a) $P=0$, which is the trivial one, or b) $x_{Q_1}=x_{P}$. 

Going to next order we encounter the last term in which also $Q_2$ appears, and again if we want a solution with $P\neq 0$ we must impose $x_{Q_2}=x_{Q_1}=x_{P}$. Therefore before solving (\ref{SPQ1})
we already know that the only consistent non-trivial solution must have all equal breaking points.

Generalizing to an arbitrary number $k$ of RSB steps we found that a consistent parametrization for the matrices $Q_1,Q_2$ and $P$ can be obtained fixing 
\begin{eqnarray}
m_i^{Q_1}=m_i^{Q_2}=m_i^P & \forall \ i=1,\ldots k\label{confinstep}
\end{eqnarray}
In the limit of infinite RSB steps the three matrices are parametrized by the functions $q_1(x),q_2(x)$ and $p(x)$ which are continuous and can eventually display constant parts (plateaus). In this case the consistent parametrization (\ref{confinstep}) corresponds to imposing that if one of the three function displays a plateau the other two functions must display a plateau of the same length located at the same position.

Looking at figure (\ref{figure}) it is clear that our parametrization is not of the type described; thus we must check its structural consistency.
To clarify what kind of problems we may encounter with such a structure for the solutions let us go back to equation (\ref{SPQ1}). 

To second order we must retain the first three terms. For $x<x_c$ we have no problems because all the three terms are varying, for $x>x_c$ $P^2$ became constant but as $Q_1^2$ has the same structure of $Q_1$ we should reasonably be able to solve the equations.

The problems arise at third order due to the presence of the last term which is proportional to $Q_2$. Let us consider the region of the second plateau of the function $q_1(x)$ in figure (\ref{figure}). In this region all the terms in (\ref{SPQ1}) that depend  on $P$ and $Q_1$ are constant because $q_1(x)$ and $p(x)$ are constant. Instead the last term $(2Q_1+Q_2)P^2$ that depends on $Q_2$ will vary in general  until the point $x_{2max}$ where the second plateau of $q_2(x)$ starts. 

Clearly if $(2Q_1+Q_2)P^2$ varies, equation (\ref{SPQ1}) cannot be verified in both the two regions $x_{1max}<x<x_{2max}$ and $x_{2max}<x<1$.

The same problem is present in the region between the ending point of the intermediate plateaus of $q_1(x)$ and $q_2(x)$, in which as one function varies the other is constant.

In a few words, the problem is that the equation for $q_1(x)$ displays terms depending on  $q_2(x)$ which for $x>x_c$ have a completely different structure. The same problem is present for the equation of $q_2(x)$ and $p(x)$.

 In the following part it is shown that the equality $p(1)=p_d$ allows us to solve these problems of structural consistency at all orders. For instance, it implies that all the terms of the form $P^{2p}Q_{1}^{q}Q_{2}^{r}$ display a plateau for  $x>x_c$ so that any dependence on the structure of $q_1(x)$ and $q_2(x)$ in this region is removed.  

The last statement can be checked by direct inspection evaluating the product $AP$ given that $P$ satisfies $p(x)=p_d$ for $x>x_c$ and $A$ is a generic Parisi matrix. It turns out that this product has the same structure of $P$, i.e. we have $(ap)(x)=(ap)_d$
 for $x>x_c$. 

In general through $p(x)=p_d$ the l.h.s. of (\ref{SPQ1}) can be recast order by order as a sum of Parisi matrices that have exactly the same structure of $Q_1$, even if they depends of $P$ and $Q_2$. The same is true for the saddle-point equation of $Q_2$ and $P$. 

These recasted terms are just the corresponding components of the covariant of $\hat{Q}$. For instance, in the equation for $Q_1$ the corresponding component of the covariant $\hat{Q}^3$ is $(Q_1^3+P^2(2Q_1+Q_2)$. The first term has the structure of $Q_1$ with two plateaus located at the same position while the second is constant for $x>x_c$; consequently $(Q_1^3+P^2(2Q_1+Q_2)$ has the structure of $Q_1$.

In other words, structural consistency is ensured because when $p(1)=p_d$ the matrix $\hat{Q}$ is separable. Writing one of its covariants  $M(\hat{Q})$ as  
\begin{equation}
M(\hat{Q})=
\left(\begin{array}{cc}
M_1(Q_1,Q_2,P) & M_{12} (Q_1,Q_2,P)
\\
M_{12}(Q_1,Q_2,P) & M_{2}(Q_1,Q_2,P)
\end{array}\right)
\end{equation}
separability for $\hat{Q}$ means that $M_1(Q_1,Q_2,P)$ has the same block structure of $Q_1$, $M_2(Q_1,Q_2,P)$ has the same block structure of $Q_2$ and $M_{12}(Q_1,Q_2,P)$ has the same block structure of $P$. 

In general, a matrix $\hat{Q}$ composed of three hierarchical matrix $Q_1,Q_2$ and $P$ is not globally separable; a sufficient condition for this is that the three matrices have the same structure, i.e. that they verify equations (\ref{confinstep}). This however is not our case. 

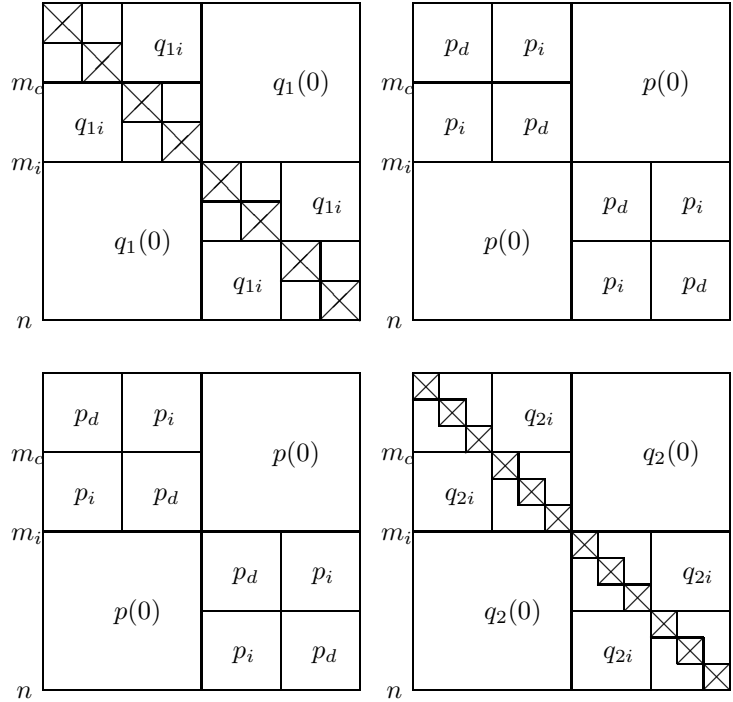
\begin{figure}[bt]
\begin{center}
\begin{picture}(260,260)(0,0)

\put(0,0){\dashbox{120}(120,120)}
\put(0,140){\dashbox{120}(120,120)}
\put(140,0){\dashbox{120}(120,120)}
\put(140,140){\dashbox{120}(120,120)}

\multiput(0,60)(60,-60){2}{\dashbox{60}(60,60)}
\multiput(0,200)(60,-60){2}{\dashbox{60}(60,60)}
\multiput(140,60)(60,-60){2}{\dashbox{60}(60,60)}
\multiput(140,200)(60,-60){2}{\dashbox{60}(60,60)}

\multiput(0,90)(30,-30){4}{\dashbox{30}(30,30)}
\multiput(0,230)(30,-30){4}{\dashbox{30}(30,30)}
\multiput(140,90)(30,-30){4}{\dashbox{30}(30,30)}
\multiput(140,230)(30,-30){4}{\dashbox{30}(30,30)}

\multiput(0,245)(15,-15){8}{\dashbox{15}(15,15)}
\put(0,260){\line(1,-1){120}}
\multiput(0,245)(15,-15){8}{\line(1,1){15}}

\multiput(140,110)(10,-10){12}{\dashbox{10}(10,10)}
\put(140,120){\line(1,-1){120}}
\multiput(140,110)(10,-10){12}{\line(1,1){10}}

\put(27,27){$p(0)$}
\put(87,87){$p(0)$}
\put(12,72){$p_i$}
\put(42,102){$p_i$}
\put(72,12){$p_i$}
\put(102,42){$p_i$}
\multiput(12,102)(30,-30){4}{$p_d$}
\put(-10,-3){$n$}
\put(-12,57){$m_i$}
\put(-12,87){$m_c$}

\put(27,167){$q_1(0)$}
\put(87,227){$q_1(0)$}
\put(12,212){$q_{1i}$}
\put(42,242){$q_{1i}$}
\put(72,152){$q_{1i}$}
\put(102,182){$q_{1i}$}
\put(-10,137){$n$}
\put(-12,197){$m_i$}
\put(-12,227){$m_c$}

\put(167,27){$q_2(0)$}
\put(227,87){$q_2(0)$}
\put(152,72){$q_{2i}$}
\put(182,102){$q_{2i}$}
\put(212,12){$q_{2i}$}
\put(242,42){$q_{2i}$}
\put(130,-3){$n$}
\put(128,57){$m_i$}
\put(128,87){$m_c$}

\put(167,167){$p(0)$}
\put(227,227){$p(0)$}
\put(152,212){$p_i$}
\put(182,242){$p_i$}
\put(212,152){$p_i$}
\put(242,182){$p_i$}
\multiput(152,242)(30,-30){4}{$p_d$}
\put(130,137){$n$}
\put(128,197){$m_i$}
\put(128,227){$m_c$}

\end{picture}
\caption{The global matrix $\hat{Q}$. The block index $m_c$ corresponds to $x_c$ in figure (\ref{figure}). For $m_c<m<n$ ($0<x<x_c$) the three matrices $Q_1,Q_2$ and $P$ have the same structure, i.e. $m_i$ is identical for the three matrices. For $1<m<m_c$ ($x_c<x<1$) the $n/m_c$ blocks of size $m_c$ in $Q_1$ have a different internal structure from the corresponding $n/m_c$ blocks in $Q_2$, while the $n/m_c$ blocks in $P$ have all their components equal to $p_d$.} \label{figure4}
\end{center}
\end{figure}

To prove that the equality $p(1)=p_d$ is a sufficient condition for the separability of $\hat{Q}$ let us take a look to figure (\ref{figure4}).
Here the matrix $\hat{Q}=\left(\begin{array}{cc}
 Q_{1} & P \\
 P^{t} & Q_{2}
 \end{array}\right)$ is sketched qualitatively according to figure (\ref{figure}).

The block index $m_c$ in figure (\ref{figure4}) corresponds to $x_c$ in figure (\ref{figure}). For $m_c<m<n$ ($0<x<x_c$) the three matrices $Q_1,Q_2$ and $P$ have the same structure i.e. the same set of $m$'s (for simplicity in (\ref{figure4}) there is only the intermediate index $m_i$ between $m_c$ and $n$). 

For $1<m<m_c$ ($x_c<x<1$) the $n/m_c$ blocks of size $m_c$ in $Q_1$ have an internal structure different from the corresponding $n/m_c$ blocks in $Q_2$, while the $n/m_c$ blocks in $P$ have all their components equal to $p_d$.

In particular the crossed blocks in $Q_1$ are identical to the blocks of the same size in the standard Parisi solution at temperature $T_1$; Those in $Q_2$ correspond to the standard solution at $T_2$.

If we consider one of the blocks of size $m_c$ in the matrix $Q_1$, say the first, we can make permutations like those of figure (\ref{figure3}) between its inner blocks leaving $Q_1$ unchanged. The invariance under this block permutations ensures that any covariant of $Q_1$ has the same structure of $Q_1$.

Now, to a block of size $m_c$ in $Q_1$ corresponds a block of size $m_c$ on the diagonal of $P$ {\em components of which are all equal to $p_d$}, this ensure that not only $Q_1$ but also the global matrix $\hat{Q}$ is left unchanged by the same block permutations.

We recall how separability follows from the invariance under these block permutations. Invariance means $\pi \hat{Q}=\hat{Q}$; so the definition of covariant, i.e. $M(\pi \hat{Q})=\pi M(\hat{Q})$ implies $\pi M(\hat{Q})=M(\hat{Q})$; these permutations exchange blocks of $M(\hat{Q})$, so that the last equality means that these blocks are equal and $M(\hat{Q})$ have the same structure of $\hat{Q}$.

In other words, the structure of a matrix like $\hat{Q}$ or $M(\hat{Q})$ is univoquely determined by the set of block permutations that leave it unchanged, and the equality $p(1)=p_d$ implies that the set of invariant block permutations of $M(\hat{Q})$ coincide with that of $\hat{Q}$. 

Instead if $p_d\neq p(1)$, in order to leave $P$ unchanged under the permutations between the internal blocks of, say, the first block of size $m_c$ of $Q_1$, we should make the same block permutations on $Q_2$. Since $Q_2$ has a different block structure from $Q_1$, it would not be left unchanged.  

\section{The SK Model with Soft-Spin Distribution}

In this section we shall apply the approach of coupled replicas to the generalization of the SK model to soft spin
distribution. This model can both be
mapped onto the SK model by a proper redefinition of the parameters
$\omega,u,v,y$\ldots appearing in the free-energy functional
(\ref{Fexp}).

In the case of continuous spins each invariant belonging to the
term $\ln$ Tr $\exp \ \left[\sum_{ab}\hat{Q}S_{a}S_{b}\right]$ in
the free-energy functional must be multiplied by a proper product
of cumulants of the soft-spin distribution. Rescaling the order
parameter by a factor $\langle S^{2}\rangle$ the mapping goes as
follows
\begin{eqnarray}
\omega \rightarrow \omega=1 &v \rightarrow v=1 & u \rightarrow
u=\frac{\langle S^{4}\rangle^{2}}{\langle S^{2}\rangle^{4}}
\nonumber
\\
y \rightarrow y=\frac{\langle S^{4}\rangle}{\langle
S^{2}\rangle^{2}} & z \rightarrow z=1 & s \rightarrow
s=\frac{\langle S^{4}\rangle}{\langle S^{2}\rangle^{2}} \nonumber
\\
t \rightarrow t=\frac{\langle S^{4}\rangle^{2}}{\langle
S^{2}\rangle^{4}}
\end{eqnarray}
Therefore the two relations (\ref{cancrel}) keep on been satisfied and to
the order we computed absence of chaos it is stable against soft
spin distribution. It is important to remember that the constant
terms in (\ref{cancrel}) belong to the  expansion of $\tau_{12}$
in powers of $\tau_{1},\tau_{2}$, which is the same as in the SK
model because of the rescaling of the order parameter. Let us
comment that for technical reasons, to consider the free energy to
the fifth order allows us to obtain a result valid up to sixth order.
We showed that the relation $p(1)=p_{d}$ implies exactly $\epsilon
=0$; this statement can be checked at all orders by using the
saddle-point equations for $p(1)$ and $p_{d}$ \cite{FPV1},
substituting the first into the second one obtains
\begin{equation}
\epsilon=2(p_{d}-p(1))\left(a_{d}-a(1)+{4 \over
3}(p^{2}(1)+p_{d}^{2}+p(1)p_{d})+2v(p_{d}^{(2)}-p^{(2)}(1))-2yp^{(2)}_{d}\right)
\end{equation}
So $\epsilon$ is equal to the difference between $p(1)$ and
$p_{d}$ multiplied by a second-order factor so that if we prove
that $p(1)=p_{d}$ to the order $n$, the free-energy difference
will be zero at least at order $n+4$.

\section{Conclusion}

The present paper's main results has been to phrase a consistent
analytical picture for absence of chaos in temperature in
mean-field spin glasses and to collect evidences through direct computation that this picture actually holds.

We made use of a model of two
coupled systems that through the replica trick can be phrased
as a variational problem with an order parameter analogous to the
standard one. We proposed a set of solutions of the model, and showed that they are intrinsically non-chaotic due to their structure. 
The actual existence of these solutions is therefore the main problem, it has been checked to fifth order in the reduced temperature for the SK model and for its generalization with soft spin distribution.

 A
particular set of cancellations in the saddle-point equations turned out to hold through subtle relations between the
parameters of the models; however, it was proven that they hold exactly at all orders in the SK model
in connection with the zero-loop
expansion of replica field-theory \cite{Kondor1}. 

Other cancellations were
checked only at finite order: they constitute the main open problem. In this respect we recall that the failure of some cancellation, even at a very high order, will destroy the whole construction leading to a dramatic change in the physical picture. The solutions will become unstable, therefore, even near the critical temperature, the function $P(q_{T1T2})$ will reduce in the thermodynamic limit to a $\delta$-function centered in zero.

We also considered the possibility that the non trivial structure we propose for the solutions be incompatible with the SP equations.
In the introduction we phrased the problem through the notion of structural consistency. It is a necessary condition to solve the SP equations. We have proven that the solutions we propose satisfy this condition at all orders in the expansion in powers of the order parameter. Furthermore, this condition imposes that the corrections one obtains considering higher powers in the expansion in the reduced temperature, change the solutions quantitatively but not qualitatively. In particular, the relation $p(x)=p_d$ for $x>x_c$ holds exactly, the approximation is in the exact value of $x_c$. Also the starting points of the intermediate plateaus of $q_1(x)$ and $q_2(x)$ at all orders are exactly equal to $x_c$.

Though these solutions were obtained in the context of coupled
systems, they are solutions of the SP equations for uncoupled systems too.
Therefore they enter the average over solutions in equation (\ref{PQ12}) implying a non trivial $P(q_{T1T2})$.

However, we warn the reader that nothing like the relation $P(q)=dx/dq$ can be written for these solutions. This relation follows from the possibility of substituting the average over solutions in (\ref{PQ}) with a summation over the indeces evaluated on the standard Parisi solution.
Instead, in the case of two or more systems this is impossible because the various solutions cannot be obtained one from the other by a permutation of the indeces \cite{FPV1}.

What is implied by our results is that $P(q_{T1T2})$ has a non-zero support from zero to a maximum value $p_{max}= q_{1\ max}+O(\tau^2)$, where $q_{1\ max}$ is the self-overlap of the states at the higher temperature. The small positive corrections to this last relation remain small at any temperatures; actually it can be proven that $p_{max}= q_{1\ max}$ holds at all temperatures at the level of accuracy of the Parisi-Toulouse approximation \cite{inprep}.

On general grounds it is 
reasonable that, like the standard Parisi solution, these solutions encode
much more information than the mere value of the free energy;
actually it seems that they lead to a quantitative description of
the bifurcation picture for the free-energy landscape discussed in
the introduction \cite{inprep}.
\\
{\bf Aknowledgements}. I deeply thank Giorgio Parisi for his constant support and his precious suggestions. I thank E. Marinari and A. Pagnani for interesting discussions. I also thank F. Ritort for a careful reading of the first version of the manuscript.

\section*{Appendix}

In this appendix we report the solutions as evaluated by solving
the saddle-point equations to fourth order in the reduced
temperatures for a given constraint $q_c$ intended to be of the
same  order of magnitude of the self-overlap of the states (i.e.
$\tau_1$ or $\tau_2$). From these expressions the maximum value of $q_c$ can be
readily obtained as the value for which the first and the second
plateau of the function corresponding to the system at the higher
temperature merge.
\begin{displaymath}
{\rm for}\ \ \ \ \ \  0\leq x \leq q_c\left( \frac{u
}{\omega}+\frac{3uv-8t\omega}{2\omega^3}(\tau_1+\tau_2)\right)
\end{displaymath}
\begin{displaymath}
q_1(x)=\left(\frac{\omega}{u}-\frac{(4\tau_1+2\tau_2)uv-(12\tau_1+4\tau_2)t\omega}{2u^2\omega}\right)x
\end{displaymath}
\begin{displaymath}
p(x)=\left(
\frac{\omega}{u}-\frac{3(\tau_1+\tau_2)uv-8(\tau_1+\tau_2)t\omega}{2u^2\omega}\right)x
\end{displaymath}
\begin{displaymath}
{\rm for}\ \ \ \ \ \  q_c\left( \frac{u
}{\omega}+\frac{3uv-8t\omega}{2\omega^3}(\tau_1+\tau_2)\right)\leq
x \leq q_c\left( \frac{2u
}{\omega}+\frac{5uv-12t\omega}{\omega^3}\tau_1+\frac{uv-4t\omega}{\omega^3}\tau_2\right)
\end{displaymath}
\begin{displaymath}
q_1(x)=\left(
1-\frac{uv-4t\omega}{2u\omega^2}(\tau_1-\tau_2)\right)q_c;\
p(x)=q_c
\end{displaymath}
\begin{displaymath}
{\rm for}\ \ \ \ \  q_c\left( \frac{2u
}{\omega}+\frac{5uv-12t\omega}{\omega^3}\tau_1+\frac{uv-4t\omega}{\omega^3}\tau_2\right)
 \leq x\leq \frac{2u\tau_1}{\omega^2}+\frac{\tau_1^2}{\omega^4}\left(
u(2u+3v+2y)+6uv-16t\omega\right)
\end{displaymath}
\begin{displaymath}
q_1(x)=\left(
\frac{\omega}{2u}-\frac{6uv-16t\omega}{4u^2\omega}\tau_1\right)x;\
p(x)=q_c
\end{displaymath}
\begin{displaymath}
{\rm for}\ \ \ \ \
\frac{2u\tau_1}{\omega^2}+\frac{\tau_1^2}{\omega^4}\left(
u(2u+3v+2y)+6uv-16t\omega\right) \leq x\leq 1
\end{displaymath}
\begin{eqnarray}
q_1(x)=\frac{\tau_1}{\omega}+\frac{\tau^2_1(2u+3v+2y)}{2\omega^3}-\frac{\tau_1^3}{24\omega^5}(-48u^2-144uv-108v^2
\nonumber
\\
-64uy-144vy-48y^2+120s\omega+208t\omega+48z\omega); \ p(x)=q_c
\nonumber
\end{eqnarray}
The function $q_2(x)$ can be obtained from $q_1(x)$ exchanging $\tau_1$ with $\tau_2$.
Given that the higher temperature is $T_1$, the maximum value of
$q_c$ for which the solutions exist works out to be
\begin{displaymath}
q_{c\ max}=q_{1\ max}+\frac{16t\omega
-6uv}{4u^2\omega^2}(\tau_2-\tau_1)\tau_1
\end{displaymath}
The above expression is valid to second order in the reduced
temperatures since the total number of matrix elements with value
$q_c$ is of order one due to the equality $p(1)=p_d$. Therefore to first order the maximum overlap it's equal to the
self-overlap $q_{EA}$ of the system at the higher temperature
with positive higher-order corrections.

This quantity is accessible to direct measurement in numerical
simulations and a comparison can be done provided that the trivial
order-parameter rescaling which has led us from (\ref{F}) to
(\ref{Fexp}) is taken into account. It's important to notice that
this rescaling doesn't change the qualitative behaviour of the
solutions, in particular the three functions $q_1(x),q_2(x)$ and
$p(x)$ remain different in the small-$x$ region.

\end{document}